\documentclass[twocolumn,prc,showpacs,superscriptaddress,amsmath,amssymb,floatfix]{revtex4}

\usepackage{graphicx}
\usepackage{epsfig}

\begin{document}

\title{Studying freeze-out and hadronization in the Landau
  hydrodynamical model}

\author{Mikl\'os Z\'et\'enyi}
\affiliation{Institute for Physics and Technology, University of Bergen, 5007
  Bergen, Norway}
\affiliation{MTA-KFKI, Research Inst.\ of Particle and Nuclear Physics, 1525
  Budapest, Hungary}

\author{L\'aszl\'o P. Csernai}
\affiliation{Institute for Physics and Technology, University of Bergen, 5007
  Bergen, Norway}
\affiliation{MTA-KFKI, Research Inst.\ of Particle and Nuclear Physics, 1525
  Budapest, Hungary}
\affiliation{Frankfurt Institute for Advanced Studies, Johann Wolfgang
  Goethe University\\
  Ruth-Moufang-Str.\ 1, 60438 Frankfurt am Main, Germany}

\date{\today}

\begin{abstract}
We study the rapidity spectra in ultra-relativistic heavy ion
collisions in the framework of the Landau hydrodynamical model. We
find that thermal smearing effects improve the agreement with
experimental results on pion rapidity spectra. We describe a simple
model of the hadronization and discuss its consequences regarding the
pion multiplicity and the increasing entropy condition.
\end{abstract}

\pacs{25.75.Nq, 24.10.Nz}

\maketitle

\section{Introduction}
The Landau hydrodynamical model \cite{Landau,Landau2} is a simple
approximate solution of the hydrodynamical equations describing the
expansion of a thin disk of static gas. The original idea was to use
hydrodynamics to describe collisions of very high energy hadrons, when
a large number of new particles are created. Today the model is also
frequently applied to the expansion of the Quark Gluon Plasma (QGP)
formed in ultra-relativistic heavy ion collisions. It provides a good
description of the shape of the rapidity distribution of pions
\cite{Wong}.

\section{Review of the Landau model}
The initial state of the Landau hydrodynamical model is a thin static
disk of thickness $\Delta$ and diameter $a$. It is assumed to be the
approximation of the overlap of two highly Lorentz contracted nuclei,
and the two geometrical sizes are assumed to be related by
\begin{equation}
  \Delta = a/\gamma,
\end{equation}
with $\gamma$ the Lorentz factor corresponding to the velocity of the
colliding nuclei in the center-of-mass frame, 
\begin{equation}
  \gamma = \frac{\sqrt{S_{NN}}}{2 m_N},
\end{equation}
where $S_{NN}$ is the invariant collision energy per colliding
nucleons of mass $m_N$.

The hydrodynamical evolution of the system is divided in two parts:
the initial longitudinal expansion (during which transverse velocities
and displacements are neglected), and the subsequent ``conic flight'',
where transverse velocities appear.

Assuming a simple equation of state (EOS) of the form 
\begin{equation}
  \label{eq:EOS}
  P = \epsilon/3,
\end{equation}
an approximate solution of the 1+1 dimensional problem of the
longitudinal expansion phase is given by Landau \cite{Landau}. This
solution is summarized (using the notation of \cite{Wong}) as
follows. The energy-density field is
\begin{equation}
  \label{eq:e_field}
  \epsilon(y_+,y_-) = \epsilon_0 \exp \left\{-\frac{4}{3}\left(y_+ +
    y_- - \sqrt{y_+ y_-} \right) \right\},
\end{equation}
where $y_{\pm}$ are logarithmic light-cone coordinates
\begin{equation}
  \label{eq:ypm}
  y_{\pm} = \ln \frac{t\pm z}{\Delta},
\end{equation}
and $\epsilon_0$ is the initial energy-density of the disk. The
rapidity field is
\begin{equation}
  \label{eq:rapi_field}
  y(t,z) = \frac{1}{2}\ln\frac{t+z}{t-z} \equiv \frac{1}{2}(y_+ - y_-).
\end{equation}
In the longitudinal expansion components of the flow four-velocity are
expressed in terms of the flow rapidity as
\begin{equation}
  \label{eq:u_y}
  u^{\mu} = (\cosh{y},0,0,\sinh{y}) = \cosh{y}(1,0,0,v_z)
\end{equation}
which can be inverted to give
\begin{equation}
  y = \frac{1}{2} \ln \frac{u^0 + u^3}{u^0 - u^3} = \tanh^{-1}{v_z}.
\end{equation}
Looking at (\ref{eq:rapi_field}) one observes that the flow rapidity
and the space-time rapidity coordinate coincide (the first expression
in (\ref{eq:rapi_field}) is exactly the definition of space-time
rapidity), similarly to the case of the Bjorken model. Because of the
coincidence of flow rapidity and space-time rapidity the space-time
coordinates of a fluid element and the components of its four-velocity
are related by
\begin{equation}
  \label{eq:u_z_t}
  u^0 = \frac{t}{\tau}, \qquad u^3 = \frac{z}{\tau},
\end{equation}
where we have introduced the proper time coordinate $\tau =
\sqrt{t^2-z^2}$.

In Refs.~\cite{Landau,Wong} an estimate of the transverse displacement
of the system is given as
\begin{equation}
  \label{eq:x_t}
  r(t,y) = \frac{t^2}{4 a \cosh^2 y}.
\end{equation}
The transition to the second phase of ``conic flight'' is assumed to
happen when this transverse displacement reaches the transverse size
of the system, $a$.  Using the relations Eq.~(\ref{eq:u_z_t}) the
condition of transition to conic flight can be written as
\begin{equation}
  \label{eq:TR_cond}
  \tau_\text{TR} = 2 a.
\end{equation}
This defines a hypersurface in space-time, which we call the transition
hypersurface. 

Although \cite{Landau,Wong} give a sketch of the hydrodynamical
evolution of the system in the second phase of conic flight, in the
calculation of observables they assume that freeze-out happens
immediately, i.e.\ not only the flight angle $\theta$ of fluid
elements is assumed to be frozen on the transition hypersurface
defined by Eq.~(\ref{eq:TR_cond}), but also the fluid is replaced by
an ensemble of non-interacting particles traveling with the same
constant velocity. Thus, the transition hypersurface is also the
freeze-out (FO) hypersurface (and the subscripts TR and FO can be
interchanged). The velocity of particles is assumed to coincide with
the velocity of the fluid element at the FO hypersurface, which means
that no smearing effect coming from the thermal distribution of
particles in the fluid is taken into account.

Equation (\ref{eq:TR_cond}) gives a relation between the time of
freeze-out and the position of the fluid element at freeze-out of the
form
\begin{equation}
  \label{eq:TR_cond2}
  t_\text{FO}(z) \equiv t_\text{TR}(z) = \sqrt{4a^2 + z^2}.
\end{equation}
This means that at the freeze-out the system can be described in terms
of functions of one variable, say $z_\text{FO}$. (Because of azimuthal
symmetry everything is independent of the azimuth angle $\phi$,
furthermore, everything is assumed to be independent of the radial
coordinate $r$ for $r<a$, and to vanish for $r>a$). Furthermore, there
is a one-to-one correspondence between $z$ and the fluid rapidity $y$
on the FO hypersurface of the form (see Eqs.~(\ref{eq:u_y}),
(\ref{eq:u_z_t}) and (\ref{eq:TR_cond}))
\begin{equation}
  z_\text{FO} = 2a \sinh{y},
\end{equation}
so instead of $z_\text{FO}$ we can use $y$ as the independent
variable. Using Eq.~(\ref{eq:TR_cond2}) the FO time can be expressed
in terms of $y$ as 
\begin{equation}
  \label{eq:tFO}
  t_{FO} = 2 a \cosh{y}.
\end{equation}

In order to get the rapidity distribution of particles we start from
the expression of the total particle number 
\begin{equation}
  \label{eq:N}
  N = \int N^{\mu}d\sigma_{\mu},
\end{equation}
with $N^{\mu} = n u^{\mu}$ the particle current ($n$ is the invariant
scalar particle density), and $d\sigma_{\mu}$ the hypersurface element
four-vector. Using cylindrical coordinates $(r,\phi,z)$ the FO
hypersurface is given by
\begin{equation}
  x^{\mu}(r,\phi,z) =
  \left(t_\text{FO}(z),r\cos\phi,r\sin\phi,z\right)
\end{equation}
and so
\begin{equation}
  \label{eq:dsigma_mu1}
  d\sigma_{\mu} = \epsilon_{\mu\nu\rho\sigma}\partial_{r}x^{\nu}
  \partial_{\phi}x^{\rho}\partial_{z}x^{\sigma} dr d\phi dz
  = \left(1,0,0,-\partial_{z}t_\text{FO}(z)\right)r dr d\phi dz.
\end{equation}
From Eq.~(\ref{eq:TR_cond2})
\begin{equation}
  \partial_{z}t_\text{FO}(z) = \tanh{y},
\end{equation}
and changing the integration variable $z$ to $y$ gives
\begin{equation}
  dz = 2a\cosh{y}dy.
\end{equation}
The $\phi$ and $r$ integrations give a factor
\begin{equation}
  \int_{0}^{2\pi}d\phi\int_{0}^{a/2}rdr \rightarrow \frac{a^2\pi}{4},
\end{equation}
so finally
\begin{equation}
  \label{eq:dsigma_mu2}
  d\sigma_{\mu} \rightarrow
  \frac{a^3\pi}{2}\left(\cosh{y},0,0,-\sinh{y}\right) dy.
\end{equation}
We now assume that the transverse component of the velocity is negligible,
so $u^0$ and $u^3$ are given by (\ref{eq:u_y}) at FO too. Therefore
\begin{equation}
  \label{eq:u_dsigma}
  u^{\mu}d\sigma_{\mu} \rightarrow \frac{a^3\pi}{2}dy.
\end{equation}
We assume that QGP is an ensemble of massless quarks and gluons, and
obeys the Stefan-Boltzmann EOS, $\epsilon \propto T^4$ (which is
consistent with (\ref{eq:EOS})). In that case 
\begin{equation}
  \label{eq:n_field}
  n \propto T^3 \propto \epsilon^{3/4} = \epsilon_{0}^{3/4}
  \exp\left\{-\left(y_{+} + y_{-} - \sqrt{y_{+} y_{-}} \right) \right\},
\end{equation}
where we have used Eq.~(\ref{eq:e_field}) in the last step. As
described in \cite{Wong}, from Eq.~(\ref{eq:ypm}) it follows that on
the FO surface $y_{\pm}\vert_\text{FO} = y_{b} \pm y$, where $y_{b} =
\ln(2a/\Delta) = \ln(\sqrt{s_{NN}}/m_N)$ is the rapidity of the
colliding nuclei in the CM frame. Therefore the exponent in
(\ref{eq:n_field}) can be written on the FO surface as $\left(y_{+} +
y_{-} - \sqrt{y_{+} y_{-}}\right)\vert_\text{FO} = 2 y_{b} -
\sqrt{y_{b}^2-y^2}$. Using this in (\ref{eq:n_field}) we get the
invariant particle number density on the FO hypersurface expressed as
a function of $y$:
\begin{equation}
\label{eq:n_y}
  n(y) \propto \exp\left\{\sqrt{y_b^2-y^2}\right\}.
\end{equation}
The rapidity distribution of particles, $dN/dy$ is obtained by
differentiating (\ref{eq:N}). Thus, making use of (\ref{eq:u_dsigma})
and (\ref{eq:n_y}) we get
\begin{equation}
  \label{eq:dN_dy}
  dN/dy \propto \exp\left\{\sqrt{y_b^2-y^2}\right\},  
\end{equation}
in agreement with \cite{Wong}.

\section{Thermal effects}
The starting point of the calculation of observables is the momentum
distribution of emitted particles, which develops at freeze-out. In
the case of sudden FO happening on a hypersurface the invariant
momentum distribution is given by the Cooper-Frye formula
\cite{Cooper_Frye}
\begin{equation}
  \label{eq:CF}
  E \frac{d^3 N}{d \mathbf{p}^3} = \int f(x,p) p^{\mu} d\sigma_{\mu},
\end{equation}
where $f(x,p)$ is the phase-space distribution of particles after
freeze-out. 

In hydrodynamics local thermal equilibrium is assumed, therefore the
$x$ dependence of $f$ is encoded in the space-time dependence of the
parameters of the thermal distributions, $T(x)$ and $u^{\mu}(x)$. (The
Stefan-Boltzmann EOS $\epsilon \propto T^4$ restricts the
considerations to baryonfree media, therefore the baryo-chemical
potential is zero.) The thermal phase-space distribution has the
general form
\begin{equation}
  \label{eq:ps_dist}
  f(x,p) = \tilde{f}\left(\frac{p^{\mu}u_{\mu}(x)}{T(x)}\right) = 
\frac{g}{(2\pi)^3}
\frac{1}{\exp\left(\frac{p^{\mu}u_{\mu}(x)}{T(x)}\right) + \xi},
\end{equation}
with $\xi$=1 for fermions, -1 for bosons and 0 for classical
particles obeying Boltzmann statistics. $g$ is a degeneracy factor
that is different for each particle species.

In the case of the Landau model parameters of the fluid taken on the
FO hypersurface depend on only one variable, which can be chosen the
fluid rapidity $y$. The expression for $T(y)$ can be obtained from the
expression of the energy density, Eq.~(\ref{eq:e_field}), assuming the
Stefan-Boltzmann EOS $\epsilon = \sigma T^4$. Following the steps
leading to the expression (\ref{eq:dN_dy}) we get the result
\begin{equation}
  \label{eq:T_y}
  T(y) = \left(\frac{\epsilon_0}{\sigma}\right)^{1/4} e^{-2y_{b}/3}
  \exp\left\{\frac{1}{3}\sqrt{y_b^2-y^2}\right\}. 
\end{equation}

The transverse velocity of fluid elements is assumed to be negligible
at FO. Components of the four-velocity of the fluid are given (in
accordance with (\ref{eq:u_y})) by
\begin{equation}
  \label{eq:u_mu}
  u^{\mu}(y) = (\cosh{y},0,0,\sinh{y}).
\end{equation}

A comment concerning the validity of the model is appropriate
here. Because of the relation $u^{\mu}u_{\mu}=1$ the inclusion of a
nonzero transverse velocity of the fluid would require the
modification of Eq.~(\ref{eq:u_y}). But the relations (\ref{eq:u_y})
are a key ingredient of the Landau model related to the separation of
longitudinal and transverse expansion of the system. Their
modification would destroy the simplicity of the model.

An estimate of the transverse velocity of fluid elements can be obtained by
differentiating (\ref{eq:x_t}). We get
\begin{equation}
  v_r = \frac{dr(t,y)}{dt} = \frac{t}{2 a \cosh^2 y}.
\end{equation}
Using Eq.~(\ref{eq:tFO}) this gives 
\begin{equation}
  v_r\vert_\text{FO} = \frac{1}{\cosh{y}}
\end{equation}
on the FO hypersurface. On the other hand Eq.~(\ref{eq:u_y}) gives
$v_z = u^3/u^0 = \tanh{y}$ for the longitudinal velocity component,
which implies 
\begin{equation}
  v^2 = v_r^2 + v_z^2 = 1,
\end{equation}
meaning that fluid elements would reach the speed if light at FO if
one included a transverse velocity with the above approximations. This
also indicates that the Landau model --- although it gives a good
description of rapidity spectra --- is not a suitable model for
describing e.g.\ the transverse momentum distribution of particles, at
least not in its simplest form.

The four-momentum of particles can be specified as
\begin{equation}
  \label{eq:p_mu}
  p^{\mu} = (m_T\cosh{y_p},p_T\cos{\phi_p},p_T\sin{\phi_p},m_T\sinh{y_p})
\end{equation}
in terms of their transverse momentum $p_T$, rapidity $y_p$ and
azimuth angle $\phi_p$, where $m_T = \sqrt{m^2+p_T^2}$ is the
transverse mass, with the mass of the particle, $m$.  From
Eqs.~(\ref{eq:u_mu}) and (\ref{eq:p_mu})
\begin{equation}
  \label{eq:pmu_umu}
  p^{\mu}u_{\mu}(x) = m_T\cosh(y-y_p).
\end{equation}

In order to calculate the rapidity distribution of emitted particles
one first has to express the invariant momentum distribution,
Eq.~(\ref{eq:CF}) in variables $p_T$, $y_p$ and $\phi_p$,
\begin{equation}
  E\frac{d^3N}{d \mathbf{p}^3} = \frac{d^3N}{p_T dp_T dy_p d\phi_p}.
\end{equation}
Integrating this, and substituting the right hand side of
(\ref{eq:CF}) we get for the rapidity spectrum
\begin{equation}
  \label{eq:rapidist}
  \frac{dN}{dy_p} = \int dp_T p_T d\phi_p E\frac{d^3N}{d \mathbf{p}^3}
  = \int dp_T p_T d\phi_p \int f(x,p) p^{\mu} d\sigma_{\mu}.
\end{equation}
Here $d\sigma^{\mu}$ is given by (\ref{eq:dsigma_mu1}). The integrand
is independent of the azimuth angles $\phi$ and $\phi_p$ corresponding
to the position of the fluid element and the particle momentum, and
the radial coordinate $r$ (see Eqs.~(\ref{eq:ps_dist}), (\ref{eq:T_y})
and (\ref{eq:pmu_umu})). This means that the effect of the
$d\sigma^{\mu}$ integration is the same as given by
(\ref{eq:dsigma_mu2}) and the integral over $\phi_p$ yields a factor
of $2\pi$. The rapidity distribution is then (also making the change of
integration variable $p_T\rightarrow m_T$)
\begin{equation}
  \label{eq:dN_dy_thermal}
  \frac{dN}{dy_p} = a^3\pi^2 \int\limits_m^\infty dm_T m_T^2
  \int\limits_{-\infty}^{\infty}dy \cosh(y-y_p)
  \tilde{f}\left(\frac{p^{\mu}u_{\mu}(y)}{T(y)}\right).
\end{equation}

\section{The hadronization phase transition}
\label{sec:hadronization}
In accordance with the Stefan-Boltzmann EOS assumed in the Landau
model the expanding quark-gluon plasma (QGP) is assumed to be an
ensemble of massless quarks, antiquarks and gluons. The EOS also
requires that the QGP is baryonfree, which is a reasonable
approximation at RHIC or LHC energies, where the particle spectra are
dominated by secondaries.

On the other hand, the post freeze-out (PFO) hadronic medium is an
ensemble of massive particles of different species. We assume that the
phase transition from QGP to hadronic matter happens at the FO
hypersurface (simultaneous FO and hadronization). Physical parameters
of the two media are related via the prescription of boundary
conditions on the FO hypersurface describing energy and momentum
conservation,
\begin{equation}
  \label{eq:emom_cons}
  \left[T^{\mu\nu}d\hat{\sigma}_{\nu}\right] = 0,
\end{equation}
where the meaning of the square bracket is $[a] =
a\vert_\text{PFO}-a\vert_\text{QGP}$. The energy-momentum tensor has the form
\begin{equation}
  \label{eq:Tmunu}
  T^{\mu\nu} = (\epsilon + P)u^{\mu}u^{\nu} - P g^{\mu\nu}  
\end{equation}
in ideal hydrodynamics. $d\hat{\sigma}_{\mu}$ is unit four-vector
normal to the FO hypersurface. In the Landau model this is given by
\begin{equation}
  \label{eq:dsigma_hat}
  d\hat{\sigma}_{\mu} =
  d{\sigma}_{\mu}/\sqrt{d{\sigma}^{\rho}d{\sigma}_{\rho}} =
  (\cosh{y},0,0,-\sinh{y}) \equiv u_{\mu}
\end{equation}
(see (\ref{eq:dsigma_mu2}) and (\ref{eq:u_mu})). The last equation
reflects a general property of the Landau model, namely, that the
four-velocity of the fluid is everywhere normal to hypersurfaces
characterized by a constant space-time rapidity $\tau$, in particular
to our FO hypersurface. This feature of the Landau model is the same
as in the Bjorken model. Making use of relation (\ref{eq:dsigma_hat})
the condition of energy-momentum conservation Eq.~(\ref{eq:emom_cons})
becomes
\begin{equation}
\label{eq:emom_cons2}
  \left[\epsilon u^{\mu}\right] = 0.
\end{equation}
The component of (\ref{eq:emom_cons2}) parallel to
$d\hat{\sigma}_{\mu}$ tells us that the invariant energy density is
constant across the FO hypersurface, $[\epsilon]=0$, while the
disappearance of the component normal to $d\hat{\sigma}_{\mu}$ implies
that the four-velocity of the fluid is unchanged during the phase
transition. 

Although the invariant energy density remains the same at FO, the EOS
changes from QGP of massless quarks and gluons to hadronic matter of
massive constituents. This causes a change of temperature even if the
FO normal, $d\sigma^{\mu}$, and the local flow velocity, $u^{\mu}$,
coincide, so that the flow does not change.

In addition to energy-momentum conservation one also has to check the
validity of the increasing entropy condition, which yields a boundary
condition of the form
\begin{equation}
\label{eq:entr_cond}
  \left[su^{\mu}\right] \ge 0,
\end{equation}
where $s$ is the invariant entropy density.

In order to reduce the numerical complexity of the system we assume
that the massive particles in the hadronic phase obey the J\"uttner
(relativistic Boltzmann) distribution. This approximation is justified
at the typical phase-transition temperatures of $T\approx
170$~MeV. Then the pressure of the medium is given by
\begin{equation}
  P(T) = \sum_{\alpha}\frac{g_{\alpha}}{\pi^2}m_{\alpha}^2 T^2
  K_2(m_{\alpha}/T),
\end{equation}
where $m_{\alpha}$ and $g_{\alpha}$ are the mass and degeneracy factor
of particle type $\alpha$, and $K_2$ is a modified Bessel function of
the second kind. The energy-density is
\begin{equation}
  \label{eq:edens_T}
  \epsilon(T) = 3 \sum_{\alpha}P_{\alpha}\left(1 +
  \frac{m_{\alpha}}{3T}\frac{K_{1\alpha}}{K_{2\alpha}}\right), 
\end{equation}
where we used the notation
\begin{equation}
  K_{i\alpha} = K_i(m_{\alpha}/T).
\end{equation}
Finally, the entropy density of the system is
\begin{equation}
  \label{eq:s_PFO}
  s(T) = \sum_{\alpha}\frac{\partial P_{\alpha}}{\partial T} =
  \sum_{\alpha}\frac{g_{\alpha}}{\pi^2} m_{\alpha}^2 \left(m_{\alpha}
  K_{1\alpha} + 4T K_{2\alpha}\right).
\end{equation}
The temperature of the medium at a given position on the FO
hypersurface is obtained by solving Eq.~(\ref{eq:edens_T})
numerically. Then the phase-space distribution function,
Eq.~(\ref{eq:ps_dist}) with $\xi=0$, is known on the FO
hypersurface, and observables can be calculated via integration --
e.g.\ the rapidity distribution as given by Eq.~(\ref{eq:rapidist}).

Now we have to give the explicit form of the EOS in the hadronic phase
by specifying the particle types present in the medium and fixing their
properties (mass and degeneracy). At ultra-relativistic energies most of
the produced particles are pions. Based on this we can model the
hadronic phase by a pion gas with $m_{\pi}=0.139$~GeV and
$g_{\pi}=3$. One expects that this model will overestimate the pion
yield because in reality some of the energy is distributed among other
hadron species. However, this model can be used as a reasonable first
approximation.

Below we briefly discuss two other possible models for the post FO medium.

At RHIC the constituent quark number scaling of flow observables has
been found \cite{NCQscaling}. This observation is consistent with simple
versions of the coalescence (or recombination) model of hadronization
(see e.g.\ \cite{NCQ_recomb}). This in turn suggests that the collective
motion of the medium is developed already at the constituent quark
level, and is basically unchanged during the process of
recombination. In practice this means that the medium can be modeled by
an ensemble of constituent quarks and antiquarks when one studies
collective observables such as momentum spectra of abundant particles.

In this approach pions need a special treatment because -- due to their
small mass -- they can not be described by the recombination of a
constituent quark and antiquarks of mass $\sim$0.3 GeV. Indeed, pions
are special in the sense that they are the Goldstone bosons associated
with the breaking of chiral symmetry, which is responsible for the
generation of hadron masses. 

Based on the above we can model the post FO hadronic medium by an
ensemble containing pions in addition to constituent $u$, $d$ and $s$
quarks and their antiquarks. We will assume the values
$m_{u,d}=0.31$~GeV and $m_s=0.5$~GeV for the constituent quark (CQ)
masses, and use the degeneracy factors $g_u = g_d = g_s = 12$ where we
have taken into account the number of colors $N_c=3$ the spin degeneracy
factor of 2, and another factor of 2 for the inclusion of antiquarks. We
assume that the explicit inclusion of pions does not affect the
degeneracy of the quarks. This is justified e.g.\ if pions are formed
earlier than other hadrons.

As another possibility, the medium can be modeled by a hadron gas
containing low mass hadron states, like pion, nucleon, kaon, $\eta$,
$\rho$, and $\omega$ mesons, and $\Delta$ particles.

\section{Pion rapidity distribution}
\label{sec:rapidist}
The BRAHMS collaboration at RHIC has measured the pion rapidity
distribution in central Au+Au collisions at $\sqrt{s_{NN}}=200$~GeV
\cite{BRAHMS_PRL,BRAHMS_JPG}. In order to obtain the same spectrum from Landau
hydrodynamics first one has to fix the parameters of the initial
condition. The radius of the gold nucleus is taken to be $r=6.98$~fm,
the Lorentz contraction factor is $\gamma=\sqrt{s_{NN}}/2 m_N =
106.6$. These fix the values of $a=2r$ and $\Delta=a/\gamma$. The
initial energy density of the system is estimated as $\epsilon_0 =
E/V$, where $V$ is the volume of the Lorentz contracted gold nucleus,
\begin{equation}
  V = \frac{4\pi}{3}r^3/\gamma.
\end{equation}
Note that the volume of the initial disk of the Landau model is 
\begin{equation}
  V_L = r^2\pi\Delta = 2r^3/\gamma = 3V/2,
\end{equation}
which means that the total energy content in the initial state of the
Landau model is a factor of 3/2 larger than the energy of the
reaction in reality.

\begin{figure}[htb]
  \begin{center}
  \includegraphics[width=8.6cm]{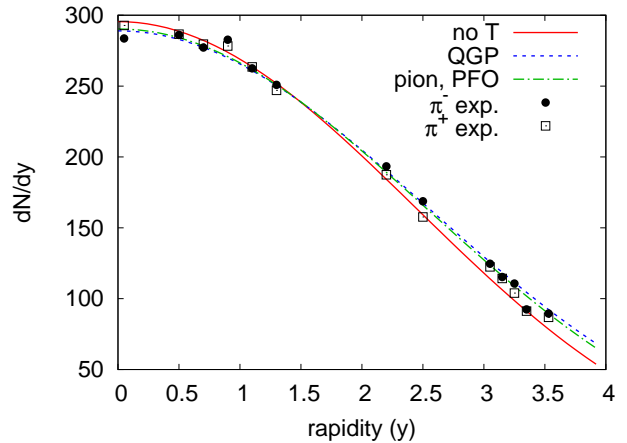}
  \caption{\label{fig:rapidist} (Color online) Rapidity distribution of
    charged pions emitted in an Au+Au collision at
    $\sqrt{s_{NN}}=200$~GeV. The experimental data points are from
    \cite{BRAHMS_PRL}. See the text for the explanation of the three
    theoretical curves.  }
  \end{center}
\end{figure}

A comparison of the pion rapidity spectra obtained from the model and
the experimental results can be seen on Fig.~\ref{fig:rapidist}. Three
theoretical curves are shown. The first one (labeled ``no T'')
corresponds to Eq.~(\ref{eq:dN_dy}). This is identical to the result of
\cite{Wong} and was obtained by neglecting thermal effects. The second
one (``QGP'') shows the result of the calculation taking into account
the thermal smearing in the QGP. This curve is the result of the
integral (\ref{eq:dN_dy_thermal}) where $\tilde{f}$ is the Fermi
distribution of massless quarks. Finally, the third one (``pion, PFO'')
is calculated from the thermal pion distribution function valid after
the phase transition, assuming a pion gas in the post FO side. In the
case of the second and third curves the integral
(\ref{eq:dN_dy_thermal}) was calculated numerically.

Looking at the shape of the obtained rapidity spectra we see that
thermal effects improve the precision of coincidence of theoretical
and experimental results. 

The author of \cite{Wong} discusses a few possible corrections to the
Landau model. One of them is related to the fact that the initial
compression of system depends on the EOS, therefore the thickness of
the initial state can deviate from the value obtained by taking into
account solely the Lorentz contraction of the nuclei. The other
possible correction comes from the observation that an arbitrary
multiplicative factor can be inserted in the condition of transition
to conic flight, Eq.~(\ref{eq:TR_cond}), since this equation expresses
only a rough equality of the transverse displacement and the
transverse system size. Both corrections lead to the modification of
the rapidity distribution, Eq.~(\ref{eq:dN_dy}), of the form
\begin{equation}
  \label{eq:dN_dy_mod}
  dN/dy \propto \exp\left\{\sqrt{(y_b+\zeta)^2-y^2}\right\},
\end{equation}
where the unknown quantity $\zeta$ can be used to improve the
agreement of the model and experiment by fitting to the experimental
data. Without questioning the validity of these arguments we note here
that some of the discrepancies of the Landau model and the
experimental data can be explained by the inclusion of thermal effects.

In the case of all three curves on Fig.~\ref{fig:rapidist} an overall
normalization factor is used following \cite{Landau,Wong} as a free
parameter and is fitted to the experimental results. This normalization
factor, $\mathcal{N}$, is defined by the equation
\begin{equation}
\left.\frac{dN}{dy_p}\right\vert_\text{normalized} =
\frac{1}{\mathcal{N}}\left.\frac{dN}{dy_p}\right\vert_\text{primary},
\end{equation}
where $\left(dN/dy_p\right)_\text{primary}$ is given by
Eq.~(\ref{eq:dN_dy_thermal}). In the case of the first two curves (``no
T'' and ``QGP'') $\mathcal{N}$ has no meaning, since the models do not
differentiate between different particle species, and give no prediction
for the total pion yield. In the third case (``pion, PFO'') on the other
hand the results are obtained from the pion phase-space distribution
function, and the value of the normalization can be used to check the
validity of the model concerning pion multiplicity.

Using the simple pion gas model for the post FO medium a normalization
factor of $\mathcal{N}=9.29$ was obtained, the primary model
overestimates the data by this factor. \footnote{A large part of this
overestimation can be attributed to the approximations used in the
Landau hydrodynamical model itself. We have already demonstrated that
the initial state of the model overestimates the total energy of the
system by a factor of 3/2. In addition, Eqs.~(\ref{eq:e_field}) and
(\ref{eq:rapi_field}) give only an approximate solution of the equations
of hydrodynamics, $\partial_{\mu} T^{\mu\nu}$, expressing local energy
and momentum conservation. Therefore, the total energy of the system is
not conserved in the model.} The total four-momentum of the system can be
calculated as
\begin{equation}
  \label{eq:ptot}
  p_\text{tot}^{\mu} = \int T^{\mu\nu} d\sigma_{\nu},
\end{equation}
where the integration is over an arbitrary spacelike hypersurface, and
the energy-momentum tensor $T^{\mu\nu}$ is given by Eq.~(\ref{eq:Tmunu})
with $\epsilon$, $P$ and $u^{\mu}$ as in (\ref{eq:e_field}),
(\ref{eq:EOS}), and (\ref{eq:u_y}), respectively. Because of the
approximations used in the model the integral Eq.~(\ref{eq:ptot}) is
{\it not independent of the choice of the hypersurface}. Specifically,
on a surface with constant proper time, $\tau$ the 0th component of
(\ref{eq:ptot}), the total energy, has the form
\begin{equation}
  \label{eq:etot}
  p_\text{tot}^{0} \equiv E_\text{tot} = \frac{a^2\pi}{4}\tau\int
  \epsilon \cosh{y} dy,
\end{equation}
where in the integration we have used steps similar to those leading to
Eq.~(\ref{eq:dsigma_mu2}). Calculating the integral Eq.~(\ref{eq:etot})
numerically on the FO hypersurface characterized by $\tau = \tau_{FO} =
2a$ (see Eq.~(\ref{eq:TR_cond})) for an Au+Au collision at
$\sqrt{s_{NN}}=200$~GeV, we get a factor of 5.16 larger value for the
total energy in the primary model than the energy of the reaction,
A$\sqrt{s_{NN}}$. An overestimation of the pion yield with the same
factor is a natural consequence.

In the experimental analysis of \cite{BRAHMS_PRL} the 5\% most central
events have been used, while the Landau model pictures an exactly
central collision. Based on a simple geometrical model describing the
colliding nuclei as spheres with uniformly distributed nucleons the
average number of participants in this experimental sample is 88\% of
the number of participants in a central collision. Assuming that the
pion multiplicity scales with the number of participants this means that
a model valid for central collisions should give a factor of $1/0.88 =
1.14$ larger result for the pion multiplicity. Thus, centrality
selection and the violation of energy conservation by the Landau
hydrodynamical model together explain a normalization factor of
$5.16\cdot 1.14 = 5.88$ in contrast to the value $\mathcal{N}=9.29$
found in the fit to the experimental pion spectra. The remaining
discrepancy can be attributed to the fact that the simple pion gas model
of the past FO medium can not account for the energy carried by other
hadrons. We can conclude that $(5.88/9.29)\cdot 100\% = 63\%$ of the
energy is carried by pions in the final state of the collision.

In \cite{BRAHMS_PRL} ratios of the full phase-space extrapolated
kaon/pion yields are given. They obtained a value of $0.173 \pm 0.017$
for the $K^+/\pi^+$ ratio and $0.143 \pm 0.014$ for the $K^-/\pi^-$. At
midrapidity the density of protons plus antiprotons, $dN_{p+\bar{p}}/dy$
is about half the density of charged kaons, $dN_{K^+ + K^-}/dy$, as can
be read off from Fig.~2 of \cite{BRAHMS_JPG}. From these one can
conclude that about 80\% of the total charged particles are pions. This
value can be consistent with our previous estimate that 63\% of total
energy is carried by pions, taking into account that -- due to their
larger mass -- kaons and protons carry more energy on average then
pions.

We also have to check the validity of the increasing entropy
condition, Eq.~(\ref{eq:entr_cond}). For that purpose we have to
specify the value of the Stefan-Boltzmann constant in the QGP EOS
$\epsilon = \sigma T^4$. Assuming a mixture of massless
quarks/antiquarks and gluons with $N_f$ quark flavors and $N_c$ colors
its value is
\begin{equation}
  \sigma = \frac{\pi^2}{15}\left(\frac{7N_f N_c}{4} + N_c^2 -1\right).
\end{equation}
Using Eq.~(\ref{eq:EOS}) we get for the entropy density in the QGP 
\begin{equation}
  s_{QGP} = \frac{\partial P}{\partial T} = \frac{4}{3}\sigma T^3.
\end{equation}
We assume a quark-gluon plasma with two quark flavors $N_f=2$, and
$N_c=3$.

\begin{figure}[htb]
  \begin{center}
  \includegraphics[width=8.6cm]{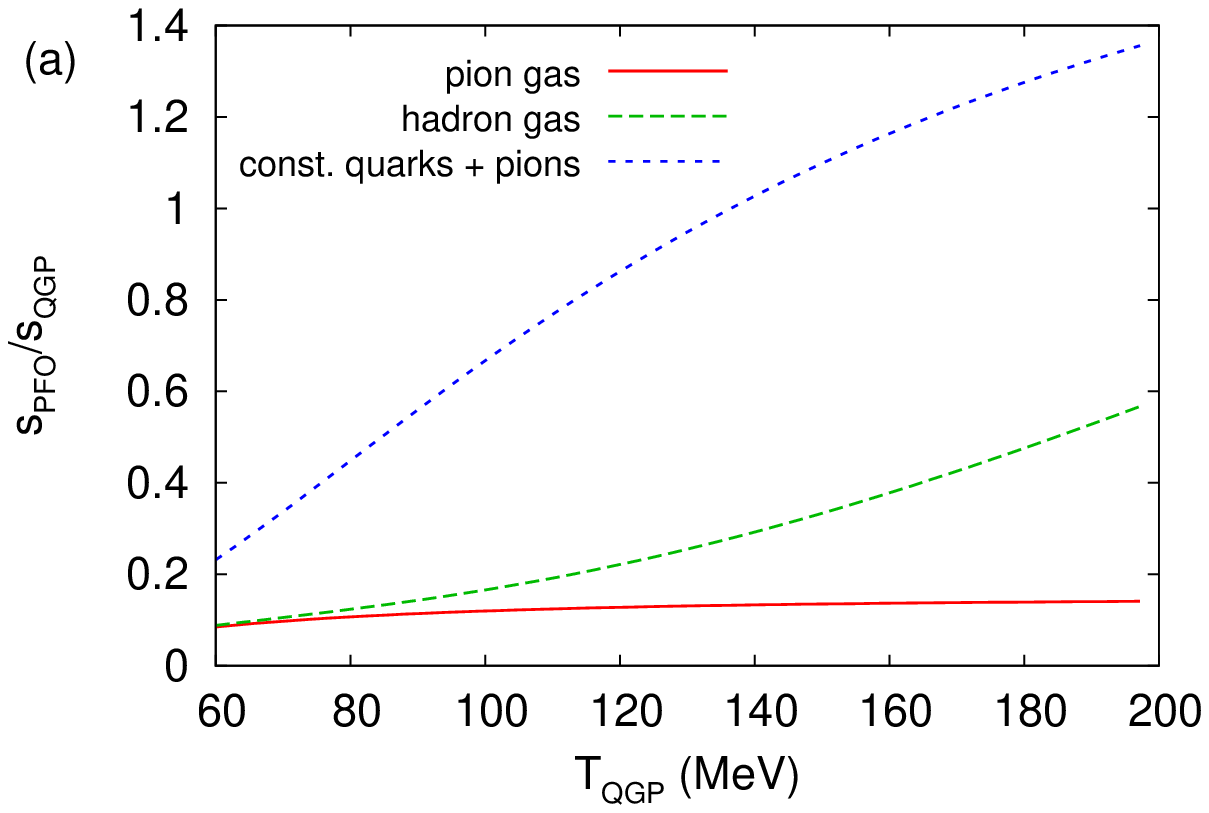}
  \includegraphics[width=8.6cm]{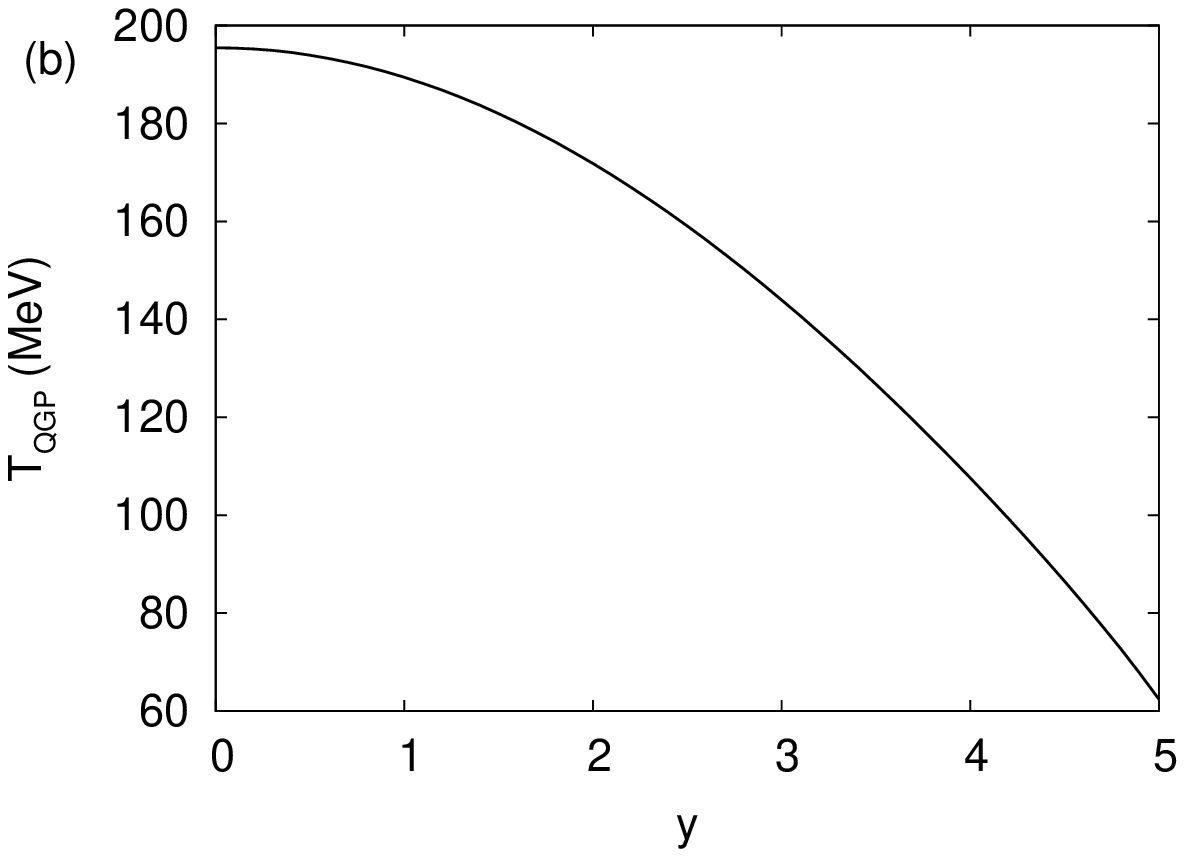}
  \caption{\label{fig:entropy} (a) (Color online) The ratio of entropy
    densities in the hadronic (or post freeze-out; PFO) state and the
    QGP state as a function of the temperature in QGP. The solid line
    represents the results obtained assuming a simple pion gas on the
    hadronic side of the phase transition. The other two curves show the
    results of the ``hadron gas'' and ``constituent quark/antiquark +
    pion'' gas models. (b) Temperature in the QGP phase at FO
    in the Landau model as a function of fluid rapidity. Parameters of
    the initial state correspond to Au+Au collisions at
    $\sqrt{s_{NN}}=200$~GeV, as described in the text.
  }
  \end{center}
\end{figure}

Because the four-velocity $u^{\mu}$ is unchanged during the phase
transition, the entropy condition (\ref{eq:entr_cond}) is equivalent to
$[s]>0$, or $s_\text{PFO}/s_\text{QGP}>1$, where $s_\text{QGP}$ and
$s_\text{PFO}$ denote the entropy density valid at the quark-gluon
plasma and post freeze-out side of the FO hypersurface,
respectively. Figure~\ref{fig:entropy}~(a) shows the ratio
$s_\text{PFO}/s_\text{QGP}$ as a function of the temperature in QGP. In
order to obtain this quantity one has to calculate the post FO
temperature in terms of the pre FO (QGP) temperature. This is done using
the QGP EOS $\epsilon = \sigma T^4$ and inverting numerically the PFO
EOS Eq.~(\ref{eq:edens_T}). Then the post FO entropy density is given by
Eq.~(\ref{eq:s_PFO}).

The continuous curve on Fig.~\ref{fig:entropy}~(a) shows that the
assumption of the simple pion gas on the hadronic side of the phase
transition violates the entropy condition, giving
$s_\text{PFO}/s_\text{QGP} \approx 0.1 - 0.15$. To understand the reason
of this we recall that the entropy carried by free massless particles in
an equilibrium ideal gas is a universal constant (3.6 for bosons, 4.2
for fermions), thus, the entropy of the system is proportional to the
number of particles. For massive particles the entropy per particle
increases but this increase is not very large for pions (see
\cite{Nonaka}). On the other hand the number of particles decreases
significantly. First of all because the degeneracy of states is reduced
drastically during the phase transition (from $4 N_F N_C=24$ for quarks
plus $N_C^2-1 = 8$ for gluons to 3 for pions). Therefore, the available
energy has to be distributed among higher energy states allowing for a
smaller number of created particles. (Also, some of the available energy
is needed to generate the pion mass.)

On Fig.~\ref{fig:entropy}~(a) we also plot the entropy density ratio for
two other possible models of the hadronic medium already mentioned at
the end of Section \ref{sec:hadronization}. Including additional hadrons
(curve ``hadron gas'' on the plot) we increase the degeneracy of the
states, and also, higher mass hadrons carry more entropy, but the
increasing entropy condition is still not fulfilled. On the other hand,
the constituent quark/antiquark + pion gas model is consistent with the
increasing entropy condition for QGP temperatures $\gtrsim 140 MeV$,
which is fulfilled for the part of the FO hypersurface where $y\lesssim
3$, as can be seen on Fig.~\ref{fig:entropy}~(b). At the edges, where
the fluid rapidity is larger, FO happens too late in the Landau
model. The increasing entropy condition favors an earlier FO at the
sides of the system.

On the other hand, the constituent quark/antiquark + pion gas model
describes an intermediate state of the system where hadrons other than
pions have not yet been formed. Exactly, the coalescence of quarks to
hadrons is the process responsible for the reduction of the number of
particles, and, thus, for the usual problems with the increasing entropy
condition.

Carrying out the calculation of the pion rapidity distribution assuming
the hadron gas model on the post FO side one finds that the shape of the
pion rapidity distribution is distorted: there is an enhancement of pion
production at large rapidities. The reason for this non-realistic pion
rapidity distribution is lying in the assumption of thermal and chemical
equilibrium in the post FO medium and the $\tau = \text{const.}$ choice
of the FO hypersurface. The FO temperature, that determines the ratio of
produced particles of different mass, is not constant along the FO
hypersurface (see Fig.~\ref{fig:entropy}~(b)). At the side of the fluid
with large fluid rapidity the temperature is low, therefore a smaller
number of large mass hadrons is created and more energy is left for pion
production. This explains the enhanced pion production at large rapidity
in this model. A key element in this effect is the presence of large
differences between the masses of the particles present in the post FO
medium.

We can conclude that the present hadronization description in
the Landau model can give a good description of the pion rapidity shape
only if the particles in the post FO medium have similar masses.

\end{document}